\documentclass{hep99}
\input{epsf}

\newcommand{\rt}{\rightarrow}
\newcommand{\ppj}{\psi^{'} \rightarrow
 \pi^+ \pi^- J/\psi}

\newcommand{\etal}{\it et al.\rm}

\newcommand{\bi}{\begin{itemize}}
\newcommand{\ei}{\end{itemize}}

\begin{document}


\title{Recent Charmonium Results from  BES}
\author{Frederick A. Harris \\
For the BES Collaboration}
%

\address{Department of Physics and Astronomy, University of Hawaii,
Honolulu, HI 96822, USA\\[3pt]
E-mail: {fah@phys.hawaii.edu}}

\abstract{This paper summarizes recent results obtained from the BES
$\psi(2S)$ event
sample, which is the world's largest.
}

\maketitle


\section{Introduction}               

The Beijing Spectrometer (BES) experiment takes data at the Beijing Electron
Positron Collider
(BEPC) which operates in the tau-charm energy range from 2 - 5 GeV.  
BES
is a conventional cylindrical magnetic detector and is 
described in detail in Ref.~\cite{bes}.
This paper describes recent preliminary results obtained from the $\psi(2S)$
data set, which is the world's largest sample.
Details can be found in the references.

\section{\boldmath $\psi(2S) \rt \pi^+ \pi^- J/\psi$ Distributions}

The dynamics of the process $\ppj$, which is the largest decay mode of
the $\psi(2S)$, can be investigated using the very clean, high
statistics sample of $\ppj, J/\psi \rt$ leptons ($\sim$ 23 K).   This reaction is
thought to occur in a two step process by the
emission of two gluons followed by hadronization to pion pairs.
Early investigation of this
decay by Mark I \cite{abrams} found that the $\pi^+ \pi^-$ mass
distribution was strongly peaked towards higher mass values, in
contrast to what is expected from phase space.  Angular distributions
strongly favored S-wave production of $\psi \pi \pi$, as well as an
S-wave decay of the $\pi \pi$ system. We observe
that the  $\cos \theta_{\pi^+}^*$ distribution, which is the cosine
of the pion angle relative to the $J/\psi$ direction in the $\pi \pi$
rest frame, is not flat as expected for S-wave decay of the dipion system.
Some D-wave
contribution is required in addition to the S-wave.

\begin{figure}[!htb]
\centerline{\epsfysize 2.6 truein
\epsfbox{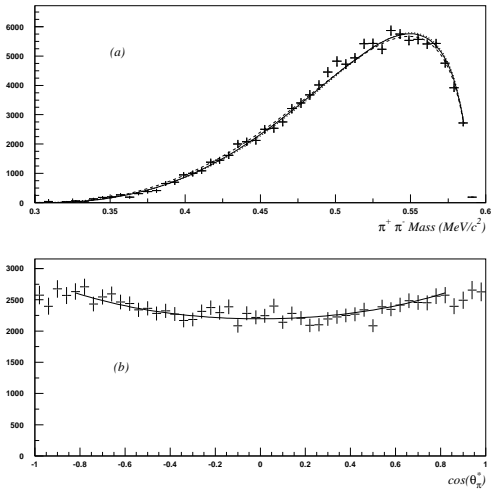}}
\caption{\label{fig:jingyun} 
Fits to 1D distributions.
{\bf (a)} $m_{\pi \pi}$
distribution (solid curve).
{\bf (b)} $\cos
\theta_{\pi}^*$ distribution.
}
\end{figure}

	One model that predicts a D-wave component is the
Novikov-Shifman model \cite{shifman}.  The pions in this process are
very low energy, so the process is nonperturbative.  This model uses
the scale anomaly and a
multipole expansion \cite{gottfried}-\cite{yan} to give an amplitude:

\begin{eqnarray}
A & \propto & \{q^2 - \kappa(\Delta M)^2(1 + \frac{2 m^2_{\pi}}{q^2}) \nonumber\\
& & + \frac{3}{2} \kappa [(\Delta M)^2 - q^2](1 - \frac{4 m^2_{\pi}}{q^2})
(\cos^2 \theta_{\pi}^* - \frac{1}{3})\}, \nonumber
\end{eqnarray}
where $q^2$ is the four-momentum squared of the dipion system,
$\Delta M = M_{\psi(2S)} - M_{J/\psi}$, and
$\kappa$ is predicted to be $\approx 0.15$ to $0.2$.
The first terms in the amplitude are the S-wave part, and the last is
the D-wave part.
Parity and charge conjugation invariance
require that the spin be even.
Fits using this amplitude are shown in Fig.~\ref{fig:jingyun}, and the
fit results are given in Table~\ref{table_kappa}.
Our result for $\kappa$ based on the $m_{\pi \pi}$
distribution is in good agreement
with that of ARGUS \cite{argus} using Mark II data for $\ppj$:  $\kappa =
0.194 \pm 0.010$ with a $\chi^2$/DOF of 38/24.

\begin{table}[!h]
\caption{Fit results for $\kappa$ for the Novikov-Shifman model.
}
\small
\begin{center}
\begin {tabular}{lcc}\hline
\label{table_kappa}
Distribution           &      $\kappa$     & $\chi^2/DOF$  \\ \hline
$m_{\pi \pi}$          & $0.186 \pm 0.003 \pm 0.010$ &  55/45     \\
$\cos \theta_{\pi}^*$   & $0.210 \pm
0.027 \pm 0.050$ &  26/40     \\
$m_{\pi \pi}$ vs $\cos \theta_{\pi}^*$     & $0.183 \pm 0.003 \pm 0.005$ &  1618/1482     \\ \hline
\end{tabular}
\end{center}
\end{table}

\begin{figure}[!htb]
\centerline{\epsfysize 2.0 truein
\epsfbox{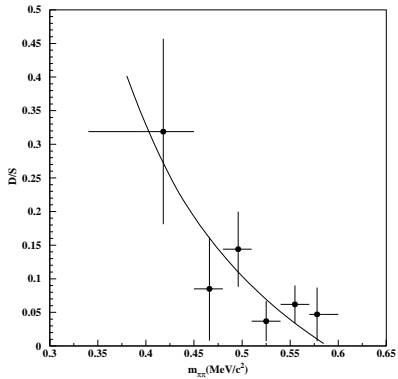}}
\caption{\label{fig:pvsm} Plot of the interference term, $\frac{D}{S}$,
versus $m_{\pi \pi}$.  The smooth curve is the prediction
of the Novikov-Shifman model for $\kappa = 0.183$.
}
\end{figure}

The amount of D wave as a function of $m_{\pi \pi}$ has been fit
using
$N(\cos \theta)  \propto  1.0 + 2(D/S)(\cos^2 \theta -1/3)
+(D/S)^2(\cos^2 \theta - 1/3)^2.$
The last term corresponds to the amount of D-wave, while the middle term
corresponds to the interference term.
The behavior of
$\frac{D}{S}$ as a function of $m_{\pi \pi}$ is shown in
Fig.~\ref{fig:pvsm}, along with the prediction of
the Novikov-Shifman model.  For more detail, see Ref.~\cite{pipijpsi}.

\section{\boldmath Hadronic $\psi(2S)$ decays}
\noindent
Both $J/\psi$ and $\psi(2S)$ decays are expected to 
proceed via $\psi \rt ggg $, with widths that are 
proportional to the square of the $c \overline{c}$ 
wave function at the origin~\cite{appel}.  
This yields the expectation that
\begin{eqnarray}
Q = \frac{B(\psi(2S) \rt X_h)}{B(J/\psi \rt X_h)} & \approx & \nonumber
\frac{B(\psi(2S) \rt e^+ e^-)}{B(J/\psi \rt e^+ e^-)} \\ \nonumber
 & = &(14.7 \pm 2.3 ) \% \nonumber
\end{eqnarray}
It was first observed by MarkII\cite{mark2} that the vector-pseudoscalar
$\rho \pi$ and $K^*\overline{K}$ channels are suppressed 
with respect to the $ 15 \% $ expectation - the
``$\rho \pi$ puzzle''.
BES finds a suppression factor of $\sim$60; this and many
other results are summarized in Ref.~ \cite{VP}.    

\begin{figure}[!htb]
\centerline{\epsfysize 2.25 truein
\epsfbox{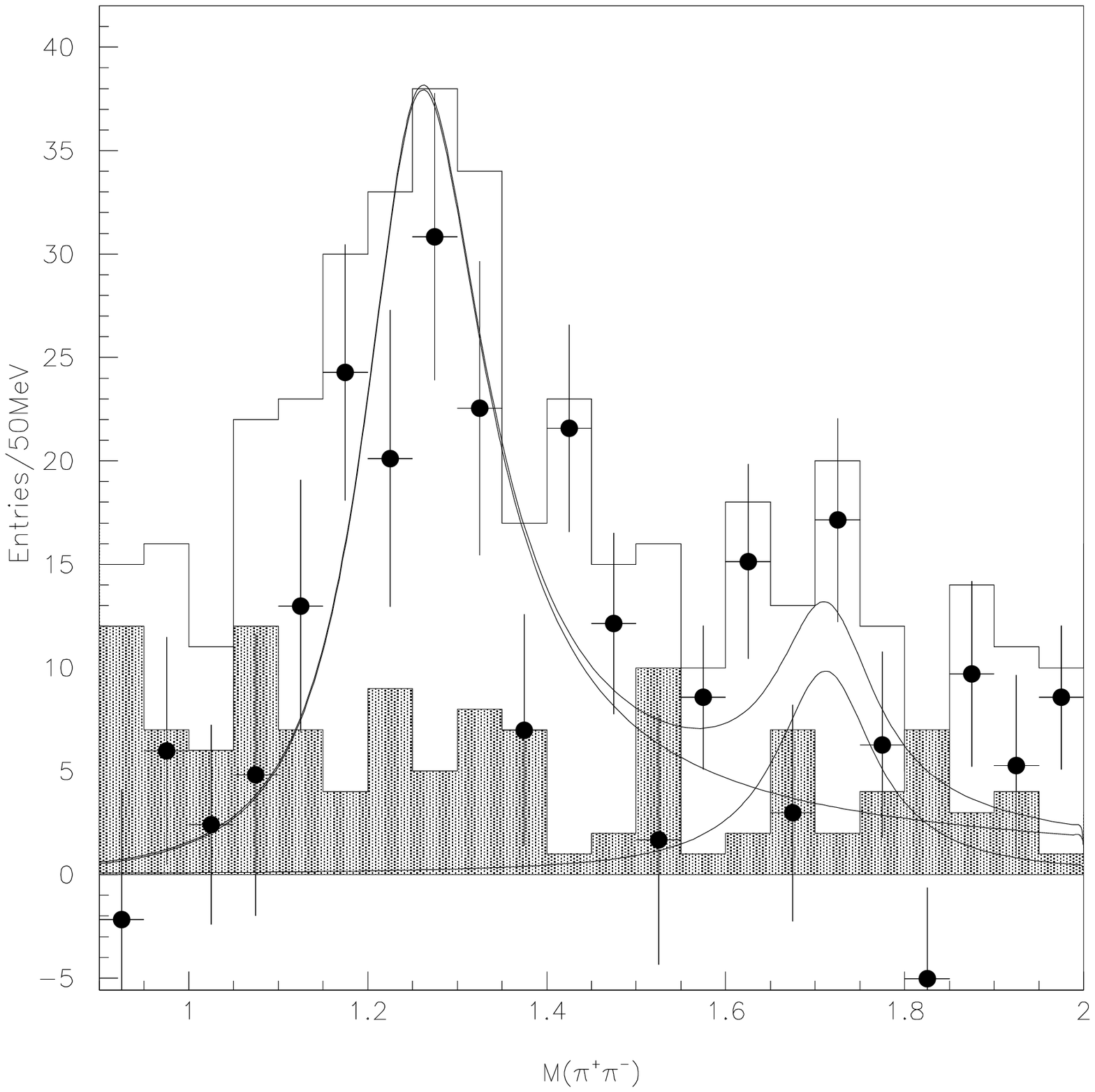}}
\caption{\label{fig:mf2pipi} 
Invariant Mass of $\pi^+ \pi^-$ in $\psi(2S)$ (histogram), $\tau$ scan
data (shaded), and the difference (points).}
\end{figure}

Here we report some recent preliminary results.
Fig.~\ref{fig:mf2pipi}
shows the $\pi^+ \pi^-$ invariant mass distribution for  $\psi(2S) \rt  \gamma
\pi^+ \pi^-$. A $f_2(1270)$ peak is seen, along with
continuum background from $e^+ e^- \rt \gamma \rho$ and $e^+ e^- \rt 
(\gamma) \mu^+ \mu^-$.
We subtract this background using our
$e^+ e^- \rt \tau^+ \tau^-$ runs \cite{tauscan}.   Fitting the peak with a D-wave Breit-Wigner
function,
we obtain
$B(\psi(2S) \rightarrow \gamma f_{2}(1270))=(2.31 \pm 0.25\pm
0.42)\times 10^{-4} $.
For  $\psi(2S)
\rt  \gamma \pi^o \pi^o$, we obtain $B(\psi(2S)\rightarrow\gamma
f_2(1270))=(3.01\pm1.12\pm1.12)\times10^{-4}$.
Combining, we find
$B(\psi(2S) \rightarrow \gamma f_{2}(1270))=(2.37 \pm 0.25 \pm
0.39)\times 10^{-4} $ and 
$
Q=\frac{B(\psi(2S) \rightarrow \gamma f_{2}(1270))}{B(J/\psi
\rightarrow \gamma f_{2}(1270))}=(17.2 \pm 1.8 \pm 2.8)\%. $

\begin{figure}[!htb]
\centerline{\epsfysize 2.25 truein
\epsfbox{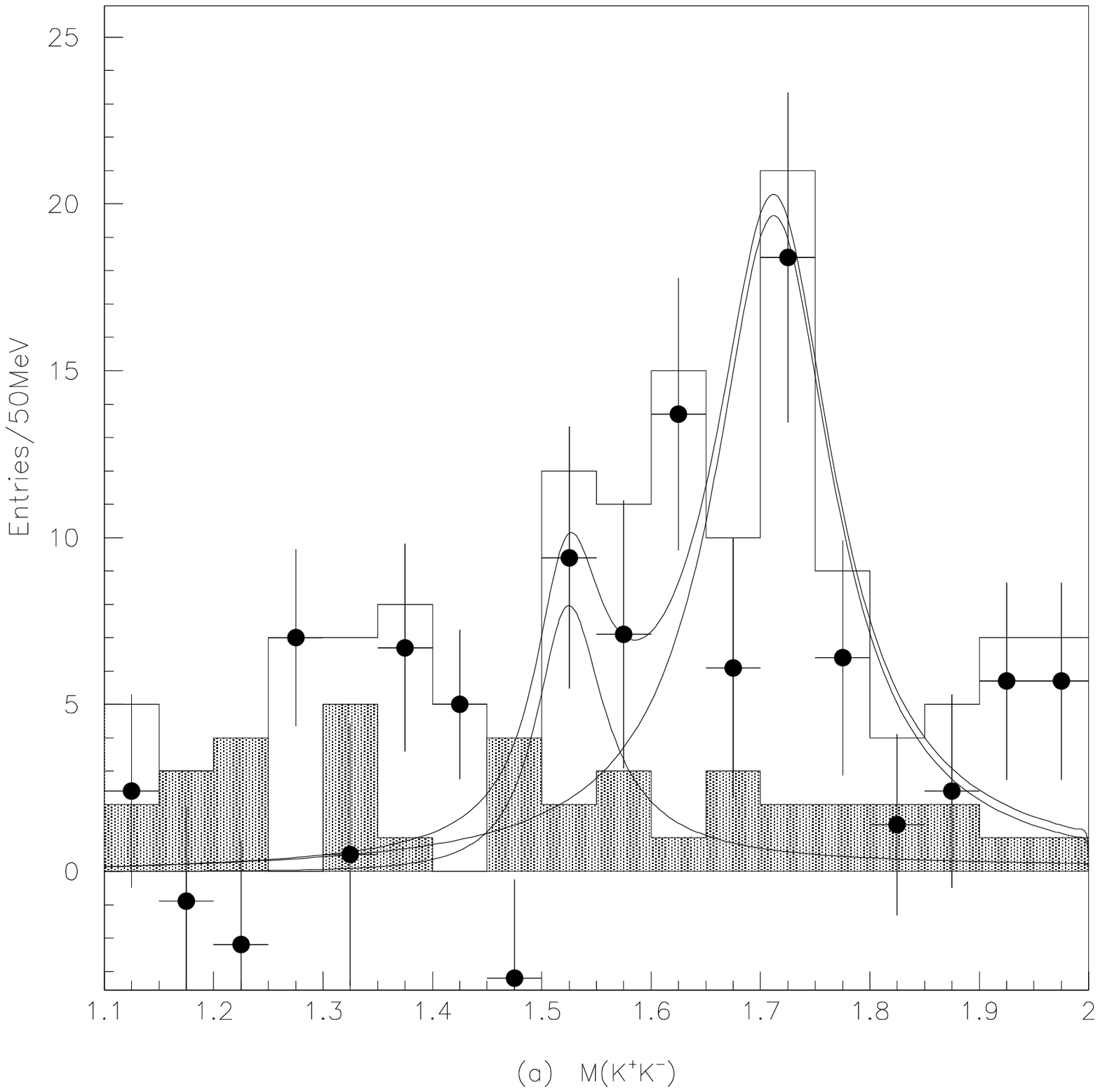}}
\caption{\label{fig:mf2kk} 
Invariant Mass of $K^+ K^-$ in $\psi(2S)$ (histogram), $\tau$ scan
data (shaded), and the difference (points).}
\end{figure}

We have also looked for $\psi(2S) \rt \gamma  K^+ K^-$. A $f_J(1710)$
signal is seen, as shown in Fig.~\ref{fig:mf2kk}.  Subtracting
continuum background and fitting with
an S-wave Breit-Wigner function, we obtain
$B(\psi(2S) \rightarrow \gamma f_J(1710))=(1.24 \pm 0.20 \pm
0.21)\times 10^{-4}$ and 
$ Q=\frac{B(\psi(2S) \rightarrow \gamma f_J(1710))}{B(J/\psi
\rightarrow \gamma f_J(1710))}=(12.8 \pm 2.6 \pm 2.7)\%.$
The ``15\% rule'' holds for $\psi(2S)$ radiative decay to
$f_2(1270)$ and
$f_J(1710)$.

For decay channels of the $\psi(2S)$ 
involving an $\omega$ or a $\phi$, we obtain the branching ratios
shown in Table~\ref{lists}. Some satisfy the 15 \% Rule, and  some do not.

\begin{table}[!h]
\begin{center}
\caption{Measurements of the $\psi$(2S) decay Branching Ratios.}
\vspace{0.4cm}
\label{lists}
\begin{tabular}{|c|c|c|c|}\hline
$\psi$(2S)${~\Rightarrow}$ & Br($\times10^{-5}$) & Q( \%) \\ \hline
$\omega\pi^+\pi^-$  &$44.7\pm5.7\pm5.4$ & $6.3\pm1.4$\\ 
$\omega K^+K^-$     & $12.6\pm4.3\pm3.9$ &$17.0\pm9.6$\\ 
$\phi\pi^+\pi^-$    & $17.6\pm2.2\pm2.4$ &$22.0\pm5.3$\\ 
$\phi K^+K^-$       & $6.10\pm1.79\pm1.51$ & $7.3\pm3.0$\\
$\phi f_0$          & $6.57\pm1.68\pm0.76$ &$20.5\pm8.1$\\ 
$\omega p\bar p$    & $6.00\pm2.34\pm1.41$ &$4.6\pm2.2$\\ 
$\phi p\bar p$      & $0.86\pm0.50\pm0.20$ &$18.9\pm13.4$\\ 
$\pi^+ \pi^- \pi^0 p\bar p$
                    & 38.5$\pm3.7\pm5.5$ & $16.7\pm7.1$\\ 
$\eta\pi^+\pi^-p\bar p$ & 25.8$\pm6.8\pm7.3$ & ? \\
$\eta p\bar p$      & 8.36$\pm4.8\pm4.8$ & $4.0\pm4.0$\\ \hline
\end{tabular}
\end{center}
\end{table}

\section{\boldmath $\chi_{c_{0,1,2}}$ and $\eta_c$ decays}
\noindent
The large sample of $\psi(2S)$ decays permits studies of
$\chi_{c0,1,2}$ decays with unprecedented precision ($ \sim 1 \times
10^6 \chi$'s).  Many BES branching fraction results may be found in
Refs.~\cite{VP},\cite{gammachi},\cite{metac}.  Many, like
$B(\chi_{c0} \rt p \bar{p})$, are measured for the
first time.

The mass and width of the  $\chi_{co}$ are not well determined in the
PDG \cite{pdg}.
Using $\chi_{c0} \rt \pi^+ \pi^-$, we find
$\Gamma_{\chi_{c0}}=14.3\pm 3.6$~MeV \cite{gammachi}, which is an
improvement on the PDG value ($14 \pm 5$ MeV) \cite{pdg} based on
two discrepant measurements.

Using many decay modes of the $\chi_{c0}$, BES determined
$M_{\chi_{c0}} = 3414.1 \pm 0.6 \pm 0.8$ MeV \cite{metac}.
The PDG value is $3417.3 \pm 2.8 $ MeV.

\begin{figure}[!htb]
\centerline{\epsfysize 2.5 truein
\epsfbox{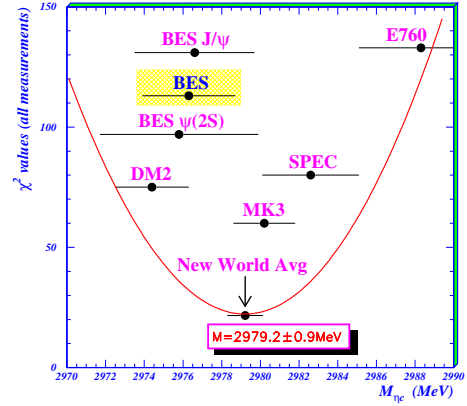}}
\caption{\label{fig:metac} 
$ M_{\eta c}$ Results}
\end{figure}

The 
mass of the $\eta_c$ is also not well determined (CL = 0.001).
We have measured the 
$\eta_c$ mass using $\eta_c \rt \pi^+ \pi^- \pi^+ \pi^-$, $\pi^+ \pi^-
K^+ K^-$, $K_s K^{\pm} \pi^{\mp}$, and $K^+ K^- K^+ K^-$,
and obtain
$M_{\eta_c} = 2975.8 \pm 3.9 \pm 1.2$ MeV \cite{metac}.
Using  the 7.8 M $J/\psi$ data, BES finds a preliminary value of  $M_{\eta_c}
= 2976.6 \pm 2.9 \pm 1.0~{\rm MeV}$, 
which agrees well with the $\psi(2S)$ result.
Combining, we obtain $M_{\eta_c} = 2976.3 \pm 2.3 \pm 0.8~{\rm MeV}$.
These and previous results are summarized in Fig.~\ref{fig:metac}.

Work supported in part by
by the National Natural
Science Foundation of China, the Chinese
Academy of Sciences,
and the Department of
Energy.

\vspace{-.2in}
\begin{thebibliography}{9}


\bibitem{bes} J.Z. Bai \etal, ~(BES Collab.), Nucl. Inst. Meth.
{\bf A344}, 319 (1994).

\bibitem{abrams} G. Abrams, Proceedings of the 1975 International
Symposium on Lepton and Photon Interactions at High Energies,
Published by the Stanford Linear Accelerator Center, 36 (1975). 

\bibitem{shifman} M. Shifman, Phys. Lett. {\bf 209}, 342 (1991).

\bibitem{gottfried} K. Gottfried, Phys. Rev. Lett. {\bf 40}, 598 (1978).

\bibitem{voloshin} M. B. Voloshin, Nucl. Phys. {\bf B 154}, 365 (1979); \\
M. Peskin, Nucl. Phys. {\bf B 156}, 365 (1979).

\bibitem{yan} T. M. Yan, Phys. Rep. {\bf 142}, 357 (1986).

\bibitem{argus} H. Albrecht \etal, ~(ARGUS Collab.), Z. Phys. {\bf C 35}, 283 (1987).

\bibitem{pipijpsi} J. Z. Bai \etal, ~submitted to PRD, 
hep-ex/9909038.

\bibitem{appel} T. Appelquist and H.D. Politzer
Phys. Rev. Lett. {\bf 34}, 43 (1975); and A. De Rujula and S. L. 
Glashow, ibid, page 46.

\bibitem{mark2} M.E.B. Franklyn \etal, ~(MarkII Collaboration), 
Phys. Rev. Lett. {\bf 51}, 963 (1983).

\bibitem{VP} F. A. Harris, {\it Recent Charmonium Results from BES}
(talk at DPF99), hep-ex/9903036.

\bibitem{tauscan} J.Z. Bai \etal, ~(BES Collab.), Phys. Rev. {\bf D53}, 20
(1996).

\bibitem{gammachi} J.Z. Bai \etal, ~(BES Collab.),
Phys. Rev. Lett. {\bf 81}, 3091 (1998). 

\bibitem{metac}  J.Z. Bai \etal, ~(BES Collab.), Phys. Rev. Lett. {\bf D60},
72001 (1999).

\bibitem{pdg} Caso \etal, ~(Particle Data Group), Eur. Phys. Jour. {\bf C3}, 1 (1998).


\end{thebibliography}
\end{document}